\documentclass[fleqn,usenatbib]{mnras}

\usepackage{newtxtext,newtxmath}

\usepackage[T1]{fontenc}
\usepackage{ae,aecompl}

\usepackage{graphicx}	
\usepackage{amsmath}	
\usepackage{amssymb}	

\title[]{The absolute magnitudes $M_J$, the binary fraction, and the binary mass ratios of M7 to M9.5 dwarfs}

\author[R. C. Laithwaite et al.]{
R. C. Laithwaite,$^{1}$
S. J. Warren,$^{1}$\thanks{E-mail: s.j.warren@imperial.ac.uk}
\\
$^{1}$Astrophysics Group, Imperial College London, Blackett Laboratory, Prince Consort Road, London SW7 2AZ\\
}
\date{Accepted 2020 September 22. Received 2020 September 9; in original form 2020 June 26}

\pubyear{2020}

\begin{document}
\label{firstpage}
\pagerange{\pageref{firstpage}--\pageref{lastpage}}
\maketitle

\begin{abstract}
We use the large homogeneous sample of late M dwarfs, M7 to M9.5, of
\citet{Ahmed19} matched to {\em Gaia} DR2, to measure the relation
between absolute magnitude and spectral type, and to infer the
multiplicity fraction of the population, and the distribution of mass
ratios in the binary systems. Binaries are identified photometrically
as overluminous sources. In order to define a sample that is unbiased
with respect to multiplicity we use distance limits that are a
function of $G-J$ colour to define a volume-complete sample of 2706
systems. The $G-J$ colours are very precise, with random errors all
less than 0.02. We measure absolute magnitudes $M_J$ that are on average 0.5
mag. brighter than previous determinations. We find evidence that the
discrepancies arise from differences in spectral types in different
samples. The measured binary fraction is $16.5\pm0.8\%$, of which $98\%$
are unresolved: both values are consistent with results of previous
studies. The distribution of excess flux in the binaries, compared to
the singles, is used to infer the mass ratio distribution $f(q)\propto
q^\gamma$, where $q=M_s/M_p$. We infer a very steep distribution over
this spectral range, with $\gamma>10$ ($99\%$ probability). This says that unresolved
ultracool M dwarf binaries reside almost exclusively in equal mass
systems, and implies that the spectral types of the unresolved
binaries match to with 0.5 spectral subtypes. The intrinsic scatter in absolute magnitude $M_J$ for ultracool M dwarfs at fixed $G-J$ colour is measured to be 0.21 mag.
\end{abstract}

\begin{keywords}
astrometry -- parallaxes -- stars: binaries, low-mass -- solar neighbourhood
\end{keywords}


\section{Introduction}

The luminosity function is a fundamental property of the population of
stars. The measurement of the field luminosity function of ultracool
dwarfs, spectral type M7 and later, has proved difficult (\citealp{Cruz2007}, \citealp{Reyle2010}, \citealp{Gagliuffi19}). The main
bottleneck in this work used to be the measurement of distances, but
with the release of {\em Gaia} DR2 \citep{2018A&A...616A...1G} this is no longer the case. Instead
the limiting factor is the time involved in spectroscopic
identification of candidates. The situation for hotter stars is
completely different. For these more luminous stars large samples of
dwarfs with accurate $G_{BP}-G_{RP}$ colours, which may be transformed
to luminosities, may be selected from the {\em Gaia} database. While
there are in fact large numbers of ultracool dwarfs detected by {\em Gaia},
at high S/N, very few have accurate {\em Gaia} colours: at $G=20$ typical uncertainties are $\sigma_G=0.01$,  but for $G_{BP}$ and $G_{RP}$ the uncertainties are 20 times greater \citep{2018A&A...616A...1G}. This means that for
measuring the luminosity function, selecting samples using {\em Gaia}
data alone is not effective, since the spectral types (equivalent to
luminosity) cannot be determined accurately.

The current state of the art for measurement of the luminosity
function, for the spectral range from M7 to L5, is the study by
\citet{Gagliuffi19} who used a volume-limited sample of 410 dwarfs
within 25 pc of the Sun. This sample is still quite moderate in size,
but increasing the size by an order of magnitude is not feasible using
the conventional route of measuring spectral types using
spectroscopy. An alternative approach is to determine spectral types
using accurate multiband photometry. \citet{Skrzypek15} developed the
{\em phototype} method, which employs 8-band $izYJHKW1W2$ photometry,
to measure accurate spectral types without spectroscopy. They applied
the method to produce a sample of 1361 L and T dwarfs brighter than
$J=17.5$ \citep{Skrzypek16}. \citet{Ahmed19} extended this work to earlier spectral types
and produced a sample of 33\,665 M7 to M9.5 dwarfs, to the same
magnitude limit. The accuracy of the spectral typing is better than
one spectral sub-type, and is competitive with spectroscopy.

The {\em phototype} method has increased the size of samples of
ultracool dwarfs with accurate spectral types by an order of
magnitude. These samples can be matched to {\em Gaia} to produce
larger volume-limited samples than existing. The {\em phototype}
samples cited above reach larger distances than {\em Gaia} for
ultracool dwarfs, primarily because they are selected in the
near-infrared. The volume of the {\em phototype} samples can therefore
be maximised by using a calibration between spectral type and
luminosity, rather than employing the {\em Gaia} distances
directly. This will be useful, for example, in investigating the local
vertical structure of the disk of the Milky Way.

The best existing calibration of the relation between absolute magnitude and spectral type
for ultracool dwarfs comes from the study of
\citet{2012ApJS..201...19D}, who used pre-{\it Gaia} ground-based parallaxes,
covering the spectral range M6 to T9. They derived polynomial
relations between absolute magnitude and spectral type, for several
photometric bands, but the number of sources remains relatively
small, $\sim10$ per subtype. The principal goal of the current paper is to use {\em Gaia}
parallaxes to determine an improved calibration between spectral type
and absolute magnitude for spectral types M7 to M9.5. This measurement
involves identifying unresolved binaries within the sample, since they
will appear as apparently overluminous sources. We therefore have expanded
the analysis to include a measurement of the multiplicty of M7 to M9.5
dwarfs. To quantify the multiplicity requires a careful consideration
of selection biases involved in matching a magnitude-limited sample to
{\it Gaia}.

Two key parameters used to describe multiplicity are the frequency of multiple systems (MF) and the companion frequency (CF).
The MF is the frequency of multiple systems as a proportion of the total number of systems.
The CF is the average number of companions per system and can exceed 100\%. There is a well-known trend of decreasing MF and CF with primary mass \citep{2013ARA&A..51..269D}. This trend continues towards the lowest masses, from early M down to the ultracool dwarfs, types $\geq$M7.  An additional important property is the distribution of mass ratios. Defining $q$ as the mass ratio $\frac{M_{\text{sec}}}{M_{\text{prim}}}\leq 1 $, it is common to characterise the distribution of mass ratios as a power law \citep{2013ARA&A..51..269D}
\begin{equation}
   f(q)\propto q^\gamma \, ,
    \label{fq}
\end{equation}
which applies over the range $0<q<1$. A large value of $\gamma$, a ``steep'' index, e.g. $\gamma>5$, implies that the components of most binaries are of approximately equal mass. 

We now summarise some of the main recent surveys for multiplicity for cool and ultracool primaries. A survey by \cite{2003ApJ...587..407C} of 39 M8.0 to L0.5 stars found a MF of $15\pm{7}\%$ for separations greater than 2.6AU. \cite{2003AJ....125.3302G} analysed 82 nearby field late M and L dwarfs using the Hubble Space Telescope and estimated a MF of $15\pm{5}\%$ for separations in the range 1.6 to 16 AU. \cite{2007ApJ...668..492A} undertook an extensive Bayesian investigation of ultracool dwarfs finding a best fit CF (approximately equal to MF, because dominated by binaries) of $\sim$20-22\%, for types M6 and later, with a wide binary frequency, defined as separation distance$>$20AU, contributing no more than 1-2\%. \cite{2008MNRAS.384..150L} found a MF of 13.6 $^{+6.5}_{-4}$ from a sample of 77 field M dwarfs from M4.5 to M6.0. \cite{2012AJ....144...64D} calculated a MF for M0-M9 dwarfs of $10.3^{+3.4}_{-2.1}\%$, for separations of $5-70$AU, based on 126 systems within $\sim$ 10 pc of the sun. A similar survey by \cite{Ward-Duong2015} found a binary fraction of $23.5\pm{3.2}\%$ for a sample of 245 K7-M6 dwarfs within 15pc over separations from 3AU to 10,000AU.
Some of these surveys provide only lower limits to the total MF, as they are not sensitive over some separation ranges. Many of these results also depend on complex corrections  for incompleteness.

The largest recent study of the multiplicity of M dwarfs is the analysis of a volume-limited survey of 1120 M dwarf primaries within 25 pc by \citet{Winters2019}. They estimated a MF of 26.8 $\pm$ 1.4$\%$ ($23.7\%$ before correction for incompleteness), nearly all in singles or binaries, with only $0.3\%$ being in triples or higher order systems. 
The MF declines with mass, and the measured MF for M dwarfs of mass $0.075-0.15 M_\odot$, uncorrected for incompleteness, is $19.8\pm3.6\%$. This corresponds to the spectral range M4 to M9. Applying the same correction factor implies a true MF of $22.4\pm4.1\%$ for this mass range, in agreement with the measurement of \citet{2007ApJ...668..492A} of $20-22\%$, cited above. The survey is limited to stellar companions of M dwarf primaries i.e. companions down to L2. There is a trend towards higher average mass ratios for lower mass primaries. This could be in part because brown dwarf companions are excluded, but they argue that including brown dwarfs would make little change to this trend. This trend towards equal mass ratios at the bottom of the main sequence agrees with the conclusion of \citet{2013ARA&A..51..269D} who argued that the flatter value of the power-law index of the distribution of mass ratios, $\gamma=1.8^{+0.4}_{-0.6}$, measured by \citet{2007ApJ...668..492A}, implies too many binaries with $q\leq 0.7$, and that the very steep value $\gamma=4.2\pm1.0$ measured by \citet{2006ApJS..166..585B} is more likely to be appropriate for very low-mass stars.

In this paper we match the large {\em phototype} sample of M7 to M9.5 stars of \citet{Ahmed19} to {\em Gaia} DR2, to determine the relation between absolute magnitude and spectral type, as well as the MF, over this spectral range. Almost all multiple systems in the sample are observed as unresolved, and are identified by virtue of being overluminous. To obtain an unbiased estimate of the MF requires selecting a volume-complete sample. For a given spectral type, the selection of appropriate upper and lower distance limits to define a sample that is unbiased with respect to multiplicity requires a careful consideration of: the flux limits of the sample of   \citet{Ahmed19}; the flux limits of the {\em Gaia} sample of sources with parallaxes; and the fact that unresolved multiple systems are brighter than single sources. Furthermore because luminosity is a strong function of spectral type, the optimal upper and lower distance limits that maximise sample size vary strongly with spectral type. We solve this problem by using the colour $G-J$ as a proxy for spectral type (or luminosity), defining upper and lower distances limits as a function of $G-J$. As we show later the $G-J$ colours are very accurate for this sample. The long baseline from $G$ to $J$ means that the colour range from M7 to M9.5 is large $\Delta(G-J)\sim1$. 

The layout of the remainder of the paper is as follows. In \S\ref{initial} we describe the sample of \citet{Ahmed19} and the matching to {\em Gaia}. In \S\ref{method} we explain the principles of selecting a volume-complete sample as a function of $G-J$ from the {\em Gaia} matched sample. In \S\ref{prepare} we provide the details of applying this procedure to the {\em Gaia} matched sample, and in \S\ref{section:model} we use a maximum-likelihood method to determine the MF, as well as the relation between $M_J$ and $G-J$, and convert this to a relation between $M_J$ and spectral type. In \S\ref{discussion} we compare our results on $M_J$ as a function of spectral type, and the measured MF, against previous determinations, and we analyse the mass ratios of the stars in the unresolved binary systems. We summarise in \S\ref{summary}. 

\section{Parent sample of M7 - M9.5 dwarfs}
\label{initial}

This study uses the homogeneous sample of M7 - M9.5 dwarfs presented by \citet{Ahmed19}. The sample comprises 33\,665 
sources over an effective area of 3\,070\,deg$^2$, with magnitude limits $13.0<J(Vega)<17.5$, on the MKO system. The sample was shown to be effectively complete except for a bias in the classification of rare peculiar blue or red objects that it is estimated affects $\sim1\%$ of sources. The sample is classified to the nearest half spectral subtype using the {\em phototype} method applied to \textit{iz} SDSS photometry  and \textit{YJHK} UKIDSS photometry. This classification was shown to be accurately calibrated to the BOSS Ultracool Dwarf (BUD) spectroscopic sample \citep{Schmidt2015}, with precision better than 0.5 subtypes rms \citep{Ahmed19}. From this sample we eliminated the 108 sources identified as peculiar, where the best template fit has $\chi^2$>20, which removes many of the subdwarfs in the sample. The sample is characterised by high S/N. For example the median and $90\%$ quantile photometric uncertainties in the J band  are 0.016 and 0.028 respectively. 

We matched this sample to the \textit{Gaia DR2}  database selecting the nearest source within a \textbf{$3\arcsec$} matching radius, and after adjusting for the average parallax zero-point shift of 0.029 mas \citep{2018A&A...616A...2L}, we calculated distances by a simple inversion of parallax i.e. $d\ (pc) = 1000 / \varpi_{adj}$. 
We note the concern of \cite{Bailer-Jones2015} that a simple inversion of \textit{Gaia} parallaxes can introduce distance inaccuracies for parallax errors $\gtrsim$ 20\%. Below we show (Figure \ref{FigVolCompRegionAll}) that our final sample has parallax errors $<$ 10\% (i.e. parallax over error $>$ 10) and therefore this approach is justified.
At this point we limited the sample to sources with positive parallaxes, and distances $<1$\,kpc. Sources at larger distances are obvious mismatches, based on the absolute magnitudes of \citet{2012ApJS..201...19D}, given the sample magnitude limit\footnote{Mismatches can occur, for example, when the M star is blended with a close neighbour of earlier spectral type. This can be a problem in {\em Gaia} because of the shorter effective wavelength of the $G$ filter, compared to the $izYJHK$ filters used in the original selection.}. This initial sample contains 14\,434 sources (`the matched parallax sample').
 
\section{Method}
\label{method}

In the section we explain the principles of selecting a large sample of ultracool M dwarfs with accurate distances, that is fully representative of the multiplicity of the population. This turns out to be a difficult problem. Our solution is to select using distance limits that are a function of $G-J$ colour. The principal difficulty is to do with the sharp decrease in luminosity in the {\em Gaia} $G$ band with spectral type, meaning that the {\em Gaia} distance limit falls sharply from M7 to M9.5. We will show later that the absolute magnitude in the G band of a M9.5 dwarf is 2.2 mag. larger than for a M7 dwarf. This means that the volume surveyed by {\em Gaia} for M9.5 dwarfs is smaller than for M7 dwarfs by a factor of 20. Therefore if we took the same distance limit for M7 dwarfs as for M9.5 dwarfs the resultant sample would be unnecessarily small. To obtain a large sample we need a distance limit that varies with spectral type.

Over the bands used {\em izYJHK} the colours are all approximately linearly related (e.g. Fig. 1 in \citet{Ahmed19}). This means that the accuracy of the spectral classifications from colours is maximised by maximising the wavelength range and minimising the photometric errors. Because of the accuracy of the {\em Gaia} $G$-band photometry and the UKIDSS $J$-band photometry, and the large wavelength range, the $G-J$ colour provides an excellent approximation to spectral type, or luminosity. We now explain how we use upper and lower distance limits that are a function of $G-J$ to select a volume-complete sample of ultra-cool M stars that is representative of the multiplicity of the population. The selection relies on knowing the relation between $M_J$ as a function of $G-J$, but since this is something we want to measure from the sample, the process is iterative. We start with an approximate measure of the relation, define the sample using it, refine the estimate, and iterate. The process converged after only one iteration.

\begin{figure}
   \centering
   \includegraphics[width=\columnwidth]{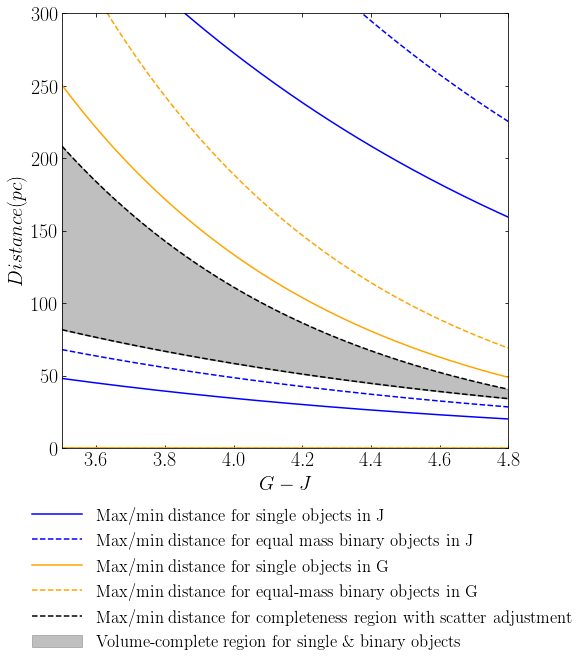}
      \caption{Illustration of method to select a volume complete sample as a function of $G-J$ colour, that includes singles and binaries, and accounts for intrinsic spread in $M_J$ at any colour.}
         \label{figExplainRegion}
\end{figure}

The selection is explained by reference to Fig. \ref{figExplainRegion}. The $G-J$ range plotted is the colour range for M7 to M9.5 stars. In this discussion we assume that the distances, from {\em Gaia}, are accurate. We discuss the effect of the parallax errors when considering the actual sample in \S\ref{prepare}. Based on \citet{Winters2019} we neglect triple and higher systems. Therefore point sources will comprise single sources, and unresolved binaries. For binaries the maximum luminosity will correspond to equal masses for the primary and secondary.  We want to ask the question over what region of this plot are all sources on the sky (single or binary) included in the matched parallax sample. Initially assume that at any particular $G-J$ colour there is negligible spread in $M_J$. Then, using the relation between $M_J$ and $G-J$, we can plot the lower and upper distance cuts for single sources at the sample magnitude limits $J=13$ and $J=17.5$ (solid blue lines). Similarly we can plot the same limits for equal mass binaries (dashed blue lines) i.e. a binary of the same apparent magnitude as a single is in fact located at a distance $\sqrt{2}$ greater. This says that the original sample of \citet{Ahmed19} is only complete within the volume defined by the lower dashed blue line and the upper solid blue line. Equal-mass binaries will be missing below the lower limit because brighter than $J=13$, and single stars will be absent above the upper limit because fainter than $J=17.5$.

We show in section \ref{GaiaMagLimit} that the {\em Gaia} DR2 sample of sources with parallaxes is highly complete down to $G=20$ i.e. essentially all sources brighter than $G=20$ will appear in the {\em Gaia} database, but the database beomes progressively incomplete fainter than $G=20$. If we know $M_J$ as a function of $G-J$, we also know $M_G$ as a function of $G-J$. In this way we can plot the distance completeness limits for single and binary sources corresponding to $G=20$. These are plotted as solid and dashed yellow lines respectively. The bright limit of {\em Gaia} is so bright that it is not relevant to defining a volume-complete sample. 

The result of this analysis shows that the matched parallax sample is complete for all singles and binaries only between the lower blue dashed line  (the distance of an unresolved equal-mass binary of $J=13$) and the solid yellow line (the distance of a single source of $G=20$). Outside these distance limits some stars of types M7 to M9.5 would have been missed either by the UKIDSS selection or by the {\em Gaia} selection. Note that the Eddington bias (the bias due to objects scattered across the magnitude limit because of photometric errors) in this sample is negligible, because of the accuracy of the $G$ and $J$ photometry. The random uncertainty for $G-J$ is less than 0.02 mag. for every source in the final sample.

The above summary is still not quite correct because there is an intrinsic scatter in $M_J$, at any $G-J$, of approximately 0.2 mag. For example an intrinsically underluminous single source at the upper distance limit described above will be fainter than $G=20$, and so may be missing from {\em Gaia}. Our final selection squeezes both the lower and upper distance limits by 0.4 mag. With an upper distance limit $d_S(G=19.6)$ corresponding to singles at $G=19.6$ and a lower distance limit $d_B(J=13.4)$ corresponding to equal mass binaries at $J=13.4$, sources of every type will be included in the matched parallax sample within this volume, regardless of absolute magnitude spread, and whether single or binary. These distance limits (they are not photometric cuts) therefore define a volume-complete sample. A sample selected within these limits will be representative of the population in terms of multiplicity, assuming the multiplicity characteristics do not vary strongly within this spectral range.

\section{Creation of the volume-complete sample}
\label{prepare}

\subsection{The {\em Gaia} $G$ magnitude limit}\label{GaiaMagLimit}

\begin{figure}
   \centering
   \includegraphics[width=\columnwidth]{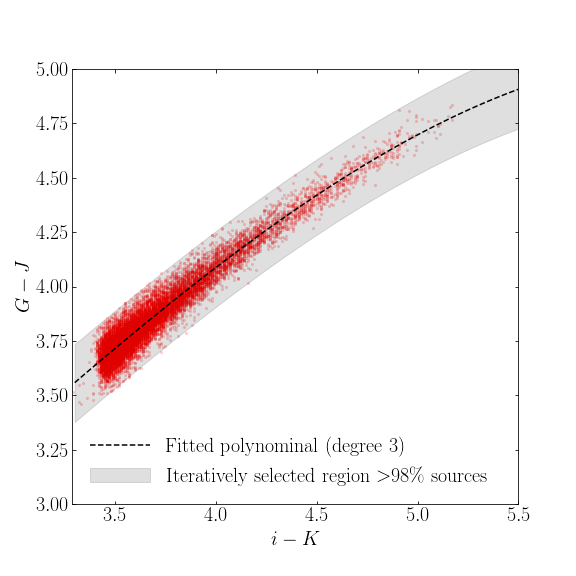}
      \caption{Colour-colour relation $G-J$ against $i-K$ for the matched sources in the sample. An iteratively derived polynomial relation of order 3 (black dotted line) is shown with rms = 0.053.  Outliers, dominated by mismatches, representing $<$ 2\% of the sample are iteratively removed.}
         \label{figColCol}
\end{figure}

The {\em Gaia} G-band limit down to which essentially all sources have a parallax in the DR2 database is a crucial quantity for establishing the upper distance limit for a volume-complete sample (as explained in \S\ref{method}). The source counts for sources with parallaxes in DR2 peaks a little fainter than $G=20$ \citep[Fig. 2,][]{2018A&A...616A...1G},  but this is only an approximation to the completeness limit. What is needed is to measure the recovery fraction as a function of $G$, for a sample with accurate $G$ photometry, and a distribution in $G$ extending, say, to $G=21$. While  this may appear difficult because the $G$ band is unique to {\em Gaia}, we can in fact create such a sample using the sample of \citet{Ahmed19} itself,  even though many of the sources are not detected in DR2. This is because the sample is homogeneous, and has high S/N, so we can predict the $G$ magnitudes from their other colours. The method to achieve this is illustrated in Fig. \ref{figColCol} which plots $G-J$ against $i-K$ for sources in the matched parallax sample. We fit a cubic polynomial to this relation, and the best fit curve  is as follows:
$$ G-J = 1.151 + 0.462\ (i-K) + 0.142\ (i-K)^2 - 0.019\ (i-K)^3\:. $$
We have then used this relation to predict the $G$ magnitudes of the full sample of \citet{Ahmed19} from the $i-K$ colours. The accuracy of the predicted $G$ magnitudes is set by the accuracy of the $i-K$ colours, the intrinsic scatter in the relation, and the gradient of the relation (the J uncertainties are very small in comparison). We find that the accuracy of the $i-K$ colours dominates over the intrinsic scatter in the relation. By this means we can establish that the predicted $G$ magnitudes are accurate to substantially better than 0.1\,mag. {\em r.m.s.}, brighter than $G=21$, which is easily sufficient for our purposes. 

\begin{figure}
   \centering
   \includegraphics[width=\columnwidth]{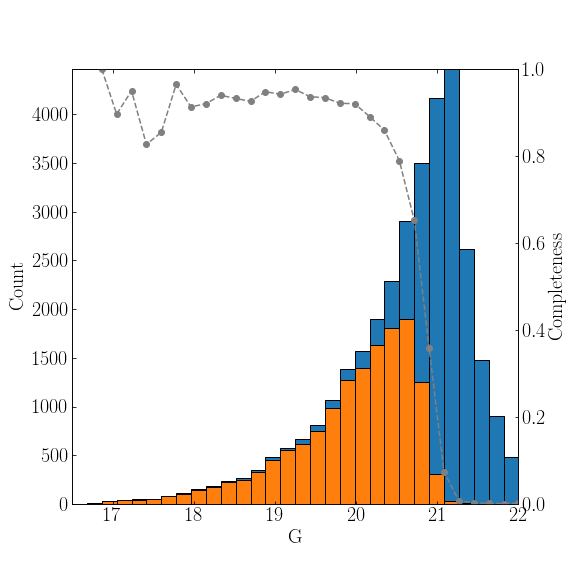}
      \caption{Histogram of sources by $G$ for the full sample of \citet{Ahmed19} (blue) and the subsample matched with GAIA (orange). The dotted line shows the ratio, which is the completeness as a function of $G$.}
         \label{FigCompletenessg}
\end{figure}

With these results we then measure the fraction of sources that matched to {\em Gaia} DR2 as a function of (predicted) $G$ magnitude. The results are plotted in Fig. \ref{FigCompletenessg}. The blue histogram is the distribution of (predicted) $G$ magnitudes of the full \citet{Ahmed19} sample, while the orange histogram represents the matched parallax sample. The completeness is given by the ratio orange/blue and is plotted as the dashed line. The completeness is very high down to $G=20$, average 94$\%$, then starts to roll over and falls rapidly beyond $G=20.3$. We therefore treat $G=20$ as the completeness limit. The small proportion of sources missing brighter than $G=20$ is associated with issues such as blending mentioned previously, flaws or diffraction spikes, and does not show a clear dependence on G magnitude. Therefore this does not affect the calculation of $M_J$ as a function of $G-J$ or the calculation of the MF. It would only affect the calculation of the space density, which we do not address here, and which could be accounted for by an adjustment to the effective area of the survey.

\begin{figure}
   \centering
   \includegraphics[width=\columnwidth]{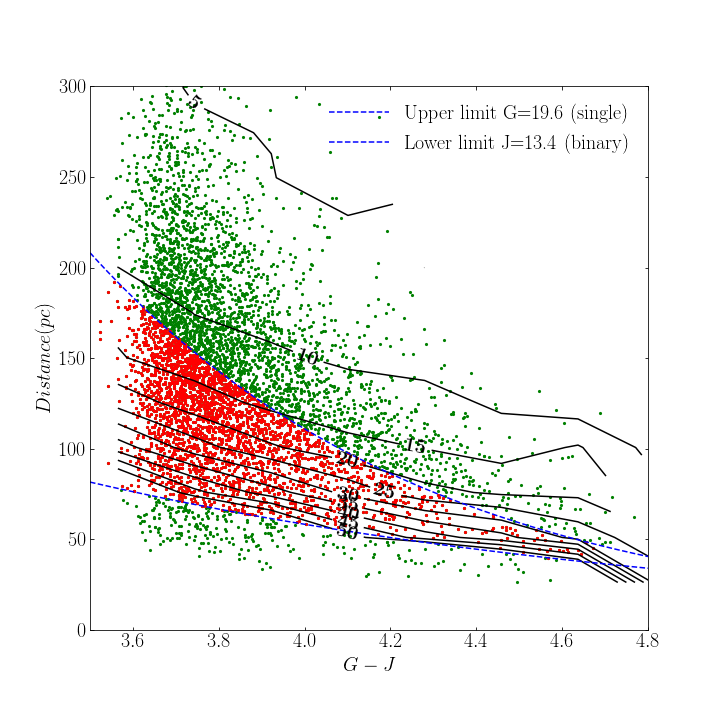}
      \caption{Plot of distance against $G-J$ for the matched parallax sample (green points) and the volume-complete subset (red points). The upper and lower distance limits, $d_S(G=19.6)$ and $d_B(J=13.4)$, are plotted as blue dashed lines. Black lines are contours of average parallax over error.}
         \label{FigVolCompRegionAll}
\end{figure}

In Fig. \ref{FigVolCompRegionAll} we plot the matched parallax sample as green points. For the sake of clarity we show only sources brighter than $G=20$. The volume-complete subset, within the distance limits $d_S(G=19.6)$ and $d_B(J=13.4)$ is plotted as red points, and comprises 2706 sources. As already explained the curves, computed from the relation between $M_J$ and $G-J$, were determined in an iterative fashion from the sample itself. The derivation of the curve is explained in Section \ref{section:model}.

We now briefly consider the possibility that any of the 2706 sources could be matched to the wrong object in {\em Gaia}, considering that the target may have moved, due to proper motion, and there may then be another nearer source within the 3\,arcsec search radius. This secondary source cannot be within 1\,arcsec of the target position as it would be blended in the original survey, and therefore the target would not have been selected. To measure the relevant surface density of {\em Gaia} sources within the survey region, we shifted the source positions by 1\,deg. in RA, and rematched to {\em Gaia}, finding only 23 matches within the annulus from 1 to 3\,arcsec radius. This quantifies the number of potential mismatches, provided the proper motion of the target is sufficiently large. But in the original matching of the 2706 sources, $68\%$ match within 1\,arcsec i.e. have small proper motion. This suggests that there could be at most about seven mismatches. Of these, some would be excluded by the distances cuts. So we may conclude that the number of mismatches is negligibly small.

\subsection{Properties of the volume-complete sample}

\begin{figure}
   \centering
   \includegraphics[width=\columnwidth]{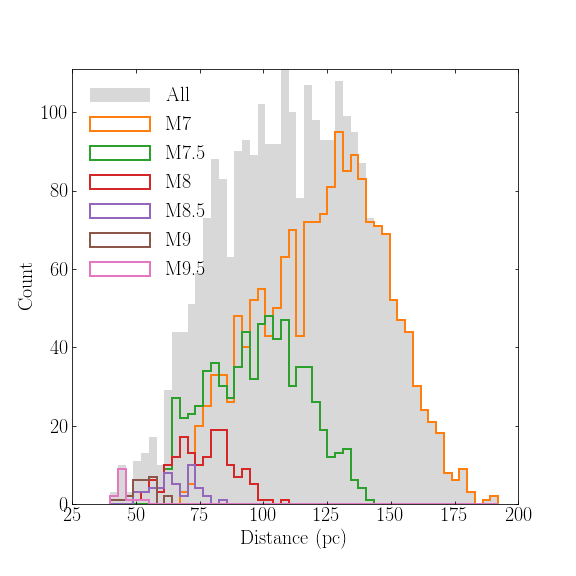}
      \caption{Histogram of volume-complete sample by distance (pc) and spectral type.}
         \label{figVolCompDist}
\end{figure}

\begin{figure}
   \centering
   \includegraphics[width=\columnwidth]{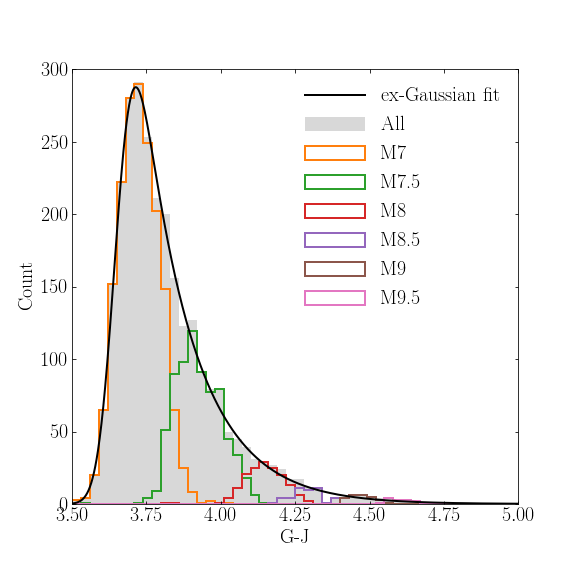}
      \caption{Histogram of volume-complete sample by G-J and spectral type. The ex-Gaussian fit is explained in Section \ref{section:modelfit}.}
         \label{figVolCompGJHist}
\end{figure}

The numbers of sources of each spectral type in the volume-complete sample of 2706 sources are provided in Table \ref{tabVolSample}. The sample is dominated by types M7 and M7.5 because the volumes sampled are much greater for these types. The distribution of distances, separated by spectral type, and for the whole sample, is plotted in Fig. \ref{figVolCompDist}. A similar plot showing the distribution of $G-J$ colours is provided in Fig. \ref{figVolCompGJHist}. This plot illustrates the fact that there is very little overlap in $G-J$ colour between the different spectral types. While this is not surprising it confirms that $G-J$ is a good proxy for luminosity, for this sample.

The distance limit cuts applied to the sample have eliminated Malmquist bias in the sample. There may be a residual bias in the sample, the Lutz-Kelker bias, depending on the accuracy of the parallaxes. In the DR2 sample the accuracy of the parallaxes depends both on the distances and the S/N, which in turn depends on the $G$ magnitude. The typical accuracy of the parallaxes is illustrated in Fig. \ref{FigVolCompRegionAll} where we plot contours of parallax-over-error for the sources in the matched parallax sample. For the volume-complete sample the 10, 50 and 90$\%$ quantiles of parallax over error are 11, 21, and 47. In the context of Lutz-Kelker bias these values are large, and according to \citet{Binney98} Lutz-Kelker bias is negligible for such large values.

\begin{table}
\centering
\begin{tabular}{lccrcc}
\hline
SpT & N & $G-J$ & $M_J$ & $M_J(DL)$ & $M_J(DL)$\\
 & & median& & mean & poly \\
\hline
\\
M7   &1737  & 3.72 &  9.92 & 10.31 & 10.65 \\ 
M7.5 & 725  & 3.92 & 10.21 & 10.82 & 10.88 \\
M8   & 159  & 4.15 & 10.56 & 10.99 & 11.06 \\
M8.5 &  46  & 4.29 & 10.78 & 11.40 & 11.19 \\ 
M9   &  25  & 4.47 & 11.05 & 11.80 & 11.31 \\
M9.5 &  14  & 4.69 & 11.25 & 11.50 & 11.41 \\
\hline
\end{tabular}
\caption{Properties of the volume-complete sample as a function of spectral type, listing: number of objects, median $G-J$ colour, absolute magnitude $M_J$ inferred from the median colour and the fitted linear relation. The final two columns list the mean $M_J$ from \citet{2012ApJS..201...19D}, and the $M_J$ computed from their polynomial relation.}
\label{tabVolSample}
\end{table}

\section{Results} \label{Modelfitting}
\label{section:model}

\subsection{Model fitting}
\label{section:modelfit}

\begin{figure}
   \centering
   \includegraphics[width=\columnwidth]{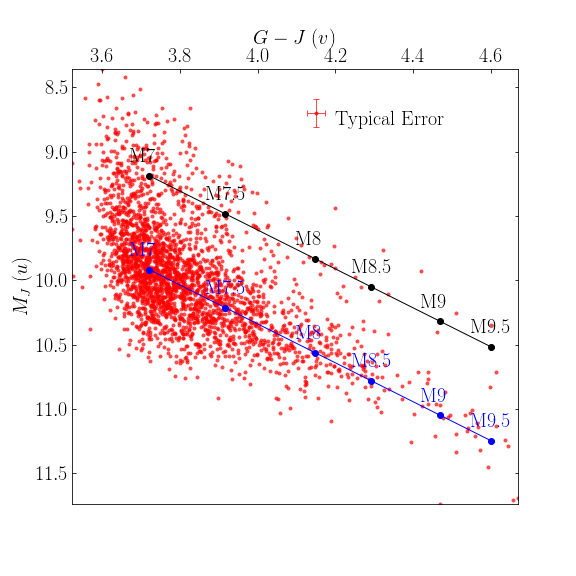}
      \caption{Colour-magnitude relation for the volume-complete sample. The blue line is the colour-magnitude relation for single stars, and the black line is the corresponding relation for binaries, as determined by the double Gaussian fit. The points on each line are plotted at the median $G-J$ colour for each spectral type.}
         \label{FigColMagSpT}
\end{figure}

The volume-complete sample is shown in Fig. \ref{FigColMagSpT}, plotting $M_J$ against $G-J$. The distribution is dominated by a band sloping down to the right, which comprises the single sources, and for which the relation between colour and absolute magnitude appears linear. Above this sequence there is a sense of a parallel sequence of overluminous sources, which are presumed to be binaries.

To model the distribution of points in this 2D space we assume a linear variation of $G-J=v$ against $M_J=u$, for the singles, parameterised by $v=au+b$. We fit a binary sequence, offset to brighter absolute magnitudes by $\Delta M$, which is a free parameter. The measured value of $\Delta M$ should then give an indication of the characteristic luminosity ratio of secondaries to primaries $L_s/L_p$.
Both the singles and binaries are characterised by a Gaussian spread in absolute magnitude, of the same standard deviation $\sigma_M$. This simplification is discussed at the end of this section.
The fraction of sources that are singles is given by $n_1$, so the fraction of binaries is $1-n_1$. The variation of the number as a function of $G-J$, plotted in Fig. \ref{figVolCompGJHist}, is parameterised by the ex-Gaussian distribution, which is the convolution of an exponential with a Gaussian, and has three parameters: mean $\mu_{col}$, standard deviation $\sigma_{col}$, and $K_{col}=\frac{1}{\sigma_{col}\lambda}$, where $\lambda$ is the exponential rate parameter.

The model therefore has 8 parameters, $a,b,\Delta M,n_1,\sigma_M,\mu_{col},\sigma_{col},K_{col}$, plus a normalisation.
The number of objects in an area element $dudv$ is $p(u,v)dudv$ where the function $p(u,v)$ is given by:
\begin{equation}
p(u,v) = p(v)\left(\frac{n_1}{\sqrt{2\pi}\sigma_M}e^{-\frac{\left(u-\frac{(v-b)}{a}\right)^2}{2\sigma_M^2}} +
    \frac{(1-n_1)}{\sqrt{2\pi}\sigma_M}e^{-\frac{\left(u+\Delta M -\frac{(v-b)}{a}\right)^2}{2\sigma_M^2}}\right) \:,
    \label{fitfunction}
\end{equation}
and $p(v)= \text{ex-Gaussian}(v| \mu_{col}, \sigma_{col}, K_{col})$. 

We adopt broad uniform priors for all the parameters, so the best fit corresponds to the maximum-likelihood solution. The likelihood assumes a Poisson point process \citep[e.g.][]{Marshall1983}. To avoid sensitivity to outliers, after obtaining the best fit, points more than 3$\sigma$ from the nearest population (i.e. binaries at high luminosities and singles at faint luminosities) were clipped, before iterating. 
The results of the fit are provided in Table \ref{tabParameters}. In Fig. \ref{figVolCompGJHist} we plot the ex-Gaussian curve as the dashed black line. The linear relation between colour and absolute magnitude for the singles, and offset by $\Delta M$ for the binaries, is plotted in Fig. \ref{FigColMagSpT}.

\begin{table}
\centering
\begin{tabular}{lr}
\hline
2D model parameter & Best fit values\\
\hline
$a$ & 0.661\\
$b$   & $-2.832$\\
$\Delta M$ & $-0.730$\\
$n_1$ &  0.855\\ 
$\sigma_M$ &  0.246\\
$\mu_{col}$   & 3.652\\ 
$\sigma_{col}$   & 0.048\\ 
$K_{col}$   & 3.585\\ 
\hline
\end{tabular}

\caption{Best fit values of model parameters from equation \ref{fitfunction}.}
\label{tabParameters}
\end{table}

\begin{figure}
\centering
\includegraphics[width=\columnwidth]{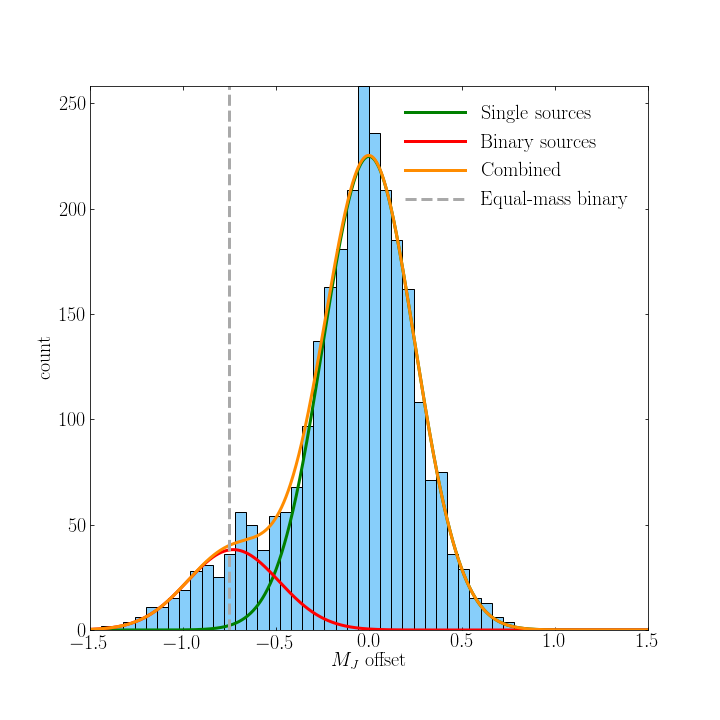}
  \caption{Histogram of the absolute magnitude offset from the linear relation for all points. The curves are the Gaussian fit for the population of singles (green) and binaries (red), as well as the sum (orange). The vertical line represents the absolute magnitude offset for a equal-mass binary (grey, dashed)}
     \label{FigModelFit}
\end{figure}

To illustrate the two populations, in Fig. \ref{FigModelFit} we plot the histogram of the horizontal distance from the linear fit, the $M_J$ offset, for all the points. Also plotted are the two Gaussians, for singles and binaries, and the sum of these, which provides a good fit to the histogram. The Gaussian fit to the binaries had no particular physical motivation, and was used simply as a tool in order to extract the colour-magnitude relation for single stars, unbiased by the presence of binaries. Nevertheless the fitted value $\Delta M=-0.73$ is striking. This has a small formal uncertainty of 0.02 and therefore is consistent with the value for binaries of equal mass of $\Delta M=-0.75$. This implies a steep distribution of mass ratios $f(q)\propto q^\gamma$ i.e. a large value of $\gamma$, and in turn provides some justification (after the event) for adopting the  same value of $\sigma_M$ for the singles and the binaries. As a check we reran the fit, with separate values of $\sigma_M$, measuring $\sigma_M=0.276\pm0.024$ for the binaries. This is slightly larger but not significantly so. The derived colour-magnitude relation was essentially identical, confirming that the fit is insensitive to the exact details of how the binaries are fit. This discussion nevertheless motivates a careful analysis of constraints on the distribution of mass ratios, which we present in section \ref{binarymass}.

\subsection{Absolute magnitudes $M_J$ for M7 to M9.5 dwarfs}

We compute the absolute magnitudes as a function of spectral type by finding the median colour of each spectral type in the final sample, and computing $M_J$ from the linear relation $M_J=(G-J-b)/a$. The colours and absolute magnitudes are provided in Table \ref{tabVolSample}. The best fit quadratic relation between $M_J$ and spectral type (SpT, where e.g. 7 represents M7) is 
\begin{equation}
    M_J = 2.249 + 1.5061 \times SpT - 0.05882 \times SpT^2 \,.
    \label{absJrelation}
\end{equation}
Also listed in Table \ref{tabVolSample} are the mean values measured by \citet{2012ApJS..201...19D} as well as the values computed from their polynomial relation.
These numbers are very different to our measured values. In Section \ref{discussion} we investigate the reasons for the differences.

The scatter in $M_J$, at any particular $G-J$, is $\sigma_M=0.246$ (Table \ref{tabParameters}). This includes a contribution from the parallax uncertainties as well as a small contribution from the uncertainties in J. Summing the variances of these individual uncertainties results in an average measurement error in $M_J$ of $0.13$. Subtracting in quadrature we calculate that the average intrinsic spread in absolute magnitude $M_J$ is $0.21$ mag. at any particular $G-J$.

\section{Discussion}
\label{discussion}

The model fit produced two interesting results. First, the derived absolute magnitudes are quite different to those provided by \citet{2012ApJS..201...19D}. Second the offset of the binary population $\Delta M=-0.73$ implies that the binary population is dominated by approximately equal mass pairs. We discuss both of these points in detail in this section. We also provide a value for the total MF for the population, by adding an estimate of the number of resolved binaries to the measured number of unresolved binaries.

\subsection{Absolute magnitudes $M_J$ compared to \citet{2012ApJS..201...19D}}

\begin{figure}
 \centering 
 \includegraphics[width=\columnwidth]{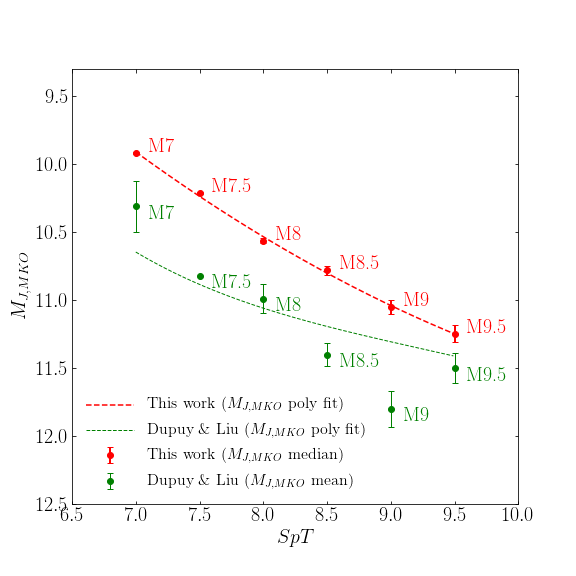} 
 \caption{Absolute magnitude as a function of spectral type from this paper (in red, upper) and from \citep{2012ApJS..201...19D} (in green, lower). The points are the median (red) or mean (green) value of $M_J(MKO)$ for that spectral type, and the curves are the polynomial relations. The mean (green) values are taken from Table 15 of  \citep{2012ApJS..201...19D} and the uncertainties are the {\em r.m.s.} values divided by $\sqrt{N}$. No error bar is plotted for M7.5 as there is only one star of this type. The formal uncertainties on the red points are mostly too small to be visible.}
 \label{figColMagDL}
\end{figure}

The discrepancy between the two measurements of absolute magnitudes is illustrated in Fig. \ref{figColMagDL}. The red points are our measured values and the dashed red curve is the quadratic fit, eqn \ref{absJrelation}. The green points are the mean values (suspected binaries removed) of the measurements of  \citet{2012ApJS..201...19D} and the green curve is their polynomial, of degree five, which is a fit applicable over the range M6 to T9. The larger scatter of the green points compared to the red points is explained by the smaller size of the sample of \citet{2012ApJS..201...19D}. There are only 25 sources M7 to M9.5 in their sample with $J(MKO)$ photometry (there are 48 with 2MASS photometry). The average offset is 0.51 mag.  

\begin{figure}
 \centering 
 \includegraphics[width=\columnwidth]{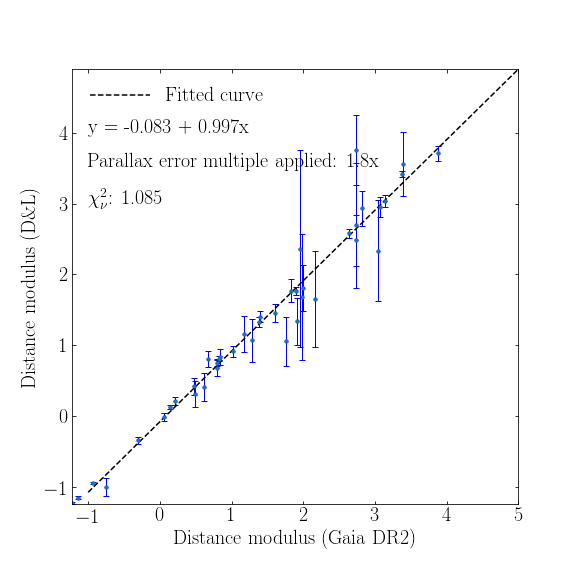} 
 \caption{Distance modulus \citep{2012ApJS..201...19D} vs. distance modulus ({\em Gaia}).}
 \label{fig:DistMod}
\end{figure}

We have confirmed that the discrepancy is not due to systematic errors in the parallaxes of \citet{2012ApJS..201...19D}. We matched their sample of M7 to M9.5 dwarfs to {\em Gaia}. In Fig. \ref{fig:DistMod} we plot their distance modulus against the distance modulus from {\em Gaia} for the 41 objects with good matches, and fit a straight line to the data, assuming that the {\em Gaia} uncertainties may be neglected in comparison. We increased the parallax uncertainties quoted by \citep{2012ApJS..201...19D} by a factor 1.8, to achieve $\chi^2_\nu\sim1$. The linear fit has a slope almost exactly unity, with small offset, indicating that any systematic errors in the parallaxes of \citep{2012ApJS..201...19D} are at the level of $<0.1$ mag. in the distance modulus and therefore cannot explain the discrepancy between our results and theirs, which are at the level of 0.5 mag.

An alternative explanation is that the discrepancy originates in the spectral types assigned to the objects. This corresponds to a horizontal shift in Fig. \ref{figColMagDL}. The average slope of the red curve over the spectral range is close to $-0.5$, meaning that there would need to be a systematic difference of one spectral type between the classifications on average to explain the differences. That is to say, for example, that a star classified M8 in the sample of \citet{2012ApJS..201...19D} would be classified M9 in \citet{Ahmed19}. This is a surprisingly large difference, but we have not found any obvious explanation. 

The spectral types in the sample of \citet{Ahmed19}, derived from $izYJHK$ colours, are very accurately calibrated to the BOSS spectroscopic sample \citep{Schmidt2015}. Comparing classifications of objects in common between the two samples, the average difference is 0.05 spectral subtypes, with a dispersion of only 0.6 subtypes \citep{Ahmed19}. Our volume-complete sample of 2706 stars includes 449 stars with BOSS classifications. These are classified to one spectral subtype. The median values of $M_J$ for single stars are 10.07, 10.53, and 11.02 for spectral types M7 (310 sources), M8 (120 sources), M9 (19 sources), after eliminating bias from the binary fraction i.e. we took the appropriate quantile assuming a binary fraction of 0.15. These values agree closely with the values in Table \ref{tabVolSample}.\footnote{The value of 10.07 for M7 is actually biased to a larger numerical value than the true value, because stars classified M7 by BOSS, but M6.5 by Ahmed and Warren, are not included here, meaning the actual agreement for M7 would be even better. This bias does not affect the M8 and M9 classifications.} Therefore the apparent difference in spectral typing is a difference between the BOSS classifications and the classifications used by \citet{2012ApJS..201...19D}. Unfortunately there are only a few stars in common between these two samples. Of stars with classifications M6.5 to M9 in \citet{2012ApJS..201...19D}, we found 10 matches in the BOSS sample. The average type difference is 0.7 subtypes, in the sense that the BOSS classifications are later. Therefore this small sample provides some support for the idea that spectral typing is the origin of the differences found in absolute magnitude. 

The classifications of \citet{2012ApJS..201...19D} come from a variety of sources and some date to a time when the classification of ultracool M dwarfs was in its infancy. The BOSS classifications are homogeneous and use spectra that are subject to uniform quality control, with good sensitivity at long wavelengths. They benefitted also from the existence of standards of L dwarfs, so the calibration is smooth through the end of the M sequence.
The BOSS methods for classification are described in \citet{West2011} and \citet{Schmidt2015}. Spectra were classified by eye, by a team, with extensive cross checks for systematics. The reason for using visual classification, despite the subjectivity, and the time involved, is that the automated classifications were judged unreliable.

While it appears that the difference between our values of $M_J$  are those of \citet{2012ApJS..201...19D} relate to differences in spectral typing it is not possible to be certain on this without going back to the original spectra used for their classifications. It is sufficient to say that our classifications are on the BOSS system, which has become the {\em de facto} standard because the sample is homogeneous and has been subject to careful checks for systematics.

\subsection{The binary mass ratio distribution}
\label{binarymass}
The offset $\Delta M=-0.73$ implies that most of the unresolved binaries comprise stars of nearly equal mass, $M_s/M_p=q\sim 1$. This in turn implies a large value of the exponent $\gamma$ of the distribution of mass ratios, $f(q)\propto q^\gamma$. This is exacerbated by the steep relation between luminosity and mass for late M dwarfs \citep{BaraffeChabrierMLR}. To see this, consider a power law relation between luminosity and mass:
\begin{equation}
    \frac{L}{L_\odot} \propto \left(\frac{M}{M_\odot}\right)^\beta \;.
    \label{MLR_UCD1}
\end{equation}
Then the relation between luminosity ratio and mass ratio of secondary to primary is
\begin{equation}
    l = \frac{L_{s}}{L_{p}} = \left(\frac{M_{s}}{M_{p}}\right)^\beta = q^\beta
    \label{MLR_UCD2}
\end{equation}
The pdf for $l$ is given by
\begin{equation}
    m(l) = \left(\frac{\gamma+1}{\beta}\right)l^{\frac{\gamma-\beta+1}{\beta}} \,.
\end{equation}
By integrating over the pdf the average value of $l$ is 
\begin{equation}
 \langle l \rangle=\frac{1+\gamma}{1+\gamma+\beta}\;.
\label{gammabeta}
\end{equation}

We now consider the consequences for $\gamma$ of the measured offset $\Delta M=-0.73$.
We define the parameter $d=\Delta M_J$ which is the horizontal distance of a binary from the linear colour-absolute magnitude relation for primaries plotted as the blue line in Fig. \ref{FigColMagSpT}. Compared to the location of the primary in this figure, addition of a secondary shifts the system in both $M_J$ and colour, such that the dependence of $d$ on $l$ is given by
\begin{equation}
    d(l) = -2.5 \log_{10}\left(\frac{(1+l)^{1+\frac{1}{a}}}{(1+l^{1+a})^\frac{1}{a}}\right) \,,
\label{deltaM}
\end{equation}
where $a$, the slope of the colour-absolute magnitude relation, is listed in Table \ref{fitfunction}. As an example an M7-M7.5 binary would have colour $G-J$ of 3.802 and $M_J$ of 9.302, and $d$ of -0.737. The binary combinations of any given primary star in Fig. \ref{FigColMagSpT} trace an arc running from the position of the single star to the equal-mass binary of the same colour. This relation means that $d$ is relatively insensitive to changes in $l$ for values of $d$ approaching the equal-mass limit.

The mass-luminosity relation becomes very steep for ultracool dwarfs. Using the values of $L/L_\odot$ and $M/M_\odot$ provided in Table 1 of \citet{BaraffeChabrierMLR}, we estimate $\beta=8$ for the spectral range M7.0 to M9.5. Here $L$ is the bolometric luminosity. The relation for luminosity measured in the $J$ band will be slightly steeper (this may be verified by using $M_K$, again from Table 1 of \citet{BaraffeChabrierMLR}, and the $J-K$ colours from \citet{Skrzypek15,Skrzypek16}). The significance of the large value of $\beta$ may be seen by inserting $\beta=8$ and $\gamma=8$ into eqn \ref{gammabeta}, which yields $\langle l \rangle=0.53$. Then the characteristic offset of the binary sequence would be 0.67 mag., from equation \ref{deltaM}. The measured offset of $0.73\pm0.02$ mag. therefore implies a large value of $\gamma$.

\begin{figure}
\centering
\includegraphics[width=\columnwidth]{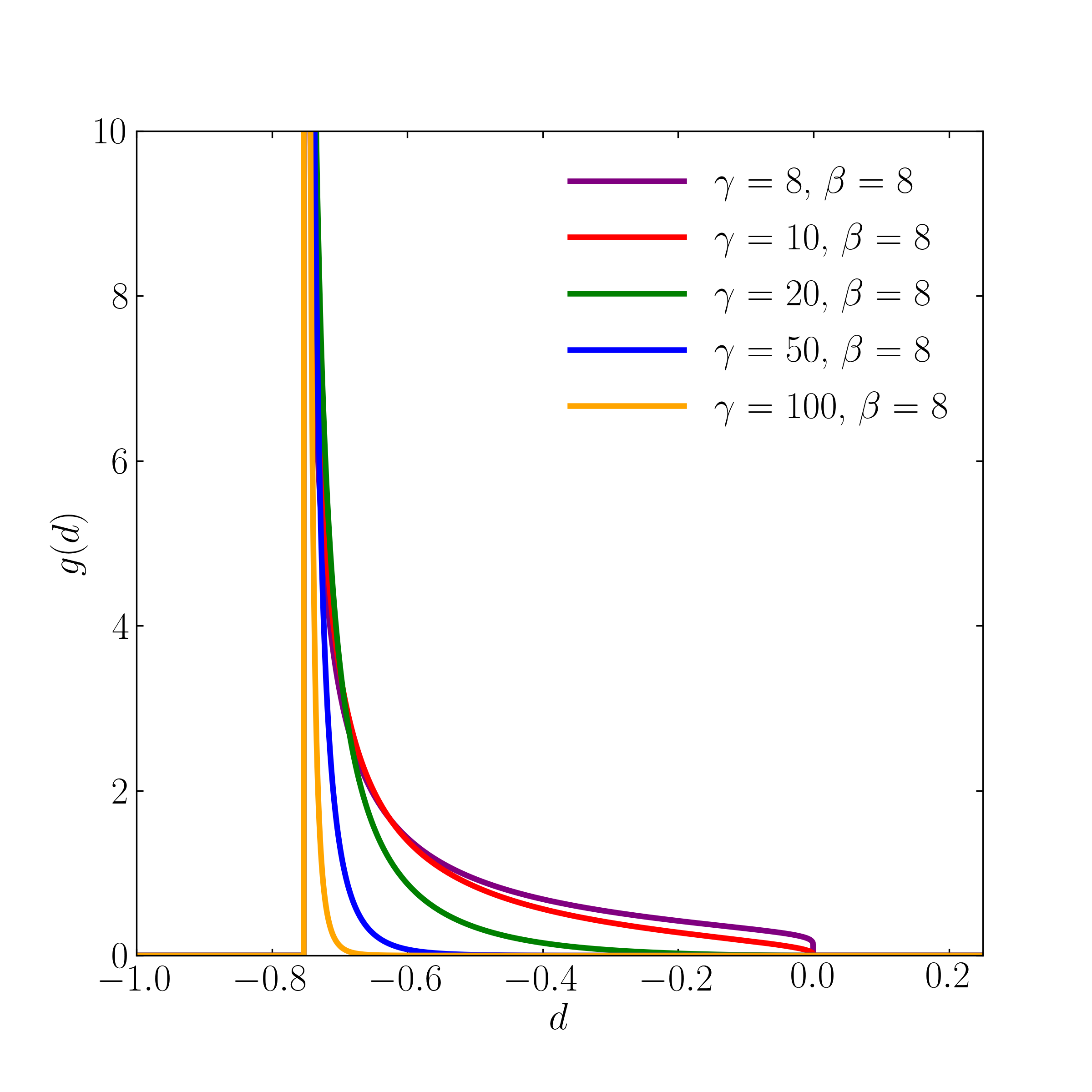}
  \caption{The pdf $g(d)$ giving the distribution of magnitude offsets for the binaries plotted for $\beta=8$ and various values of $\gamma$.}
 \label{figBinaryDist}
\end{figure}

It is possible to actually measure $\gamma$ by fitting to the distribution of offsets in absolute magnitude i.e. the data plotted as a histogram in Fig. \ref{FigModelFit}, as follows. Suppose firstly that there is no intrinsic spread in absolute magnitude of stars of a particular mass. Then the pdf of the distribution of stars in the parameter $d$ is
\begin{equation}
p(d)=(1-n_1)g(d)+n_1\delta(d) \,,
\end{equation}
where $\delta(d)$ is the Dirac delta function, $n_1$ is the fraction of systems that are single stars (as before), and $g(d)$ is given by
\begin{equation}
    g(d(l)) = \frac{m(l)}{\frac{dd}{dl}} \,.
\end{equation}
Differentiating equation \ref{deltaM} gives
\begin{equation}
    \frac{dd}{dl} = -\frac{2.5}{\ln(10)}\left(\frac{(1+\frac{1}{a})}{(1+l)}
    - \frac{(1+a) l^a}{a (1+l^{1+a})}\right) \,.
\end{equation}
The function $g(d)$ is the pdf for the distribution of binaries over the parameter $d$, as a function of $\gamma$ and $\beta$, and is defined within the range $-2.5\log_{10}(2)<d<0$, equivalent to $1>l>0$. Examples of $g(d)$ for different combinations of $\gamma$ and $\beta$ are plotted in Figure \ref{figBinaryDist}. 

To allow for an intrinsic dispersion in absolute magnitude we convolve $p(d)$ with a Gaussian of standard deviation $\sigma_M$, then multiply by a normalisation $N$ and fit to the data, computing the likelihood in the same manner as before. Fixing $\beta$, the free parameters are $\gamma, n_1, N, \sigma_M$. Note that as $\gamma$ becomes very large $p(d)\rightarrow(1-n_1)\delta(d+0.75)+n_1\delta(d)$ and the model becomes a double Gaussian, with the binaries offset by 0.75 mag.

\begin{figure}
\centering
\includegraphics[width=\columnwidth]{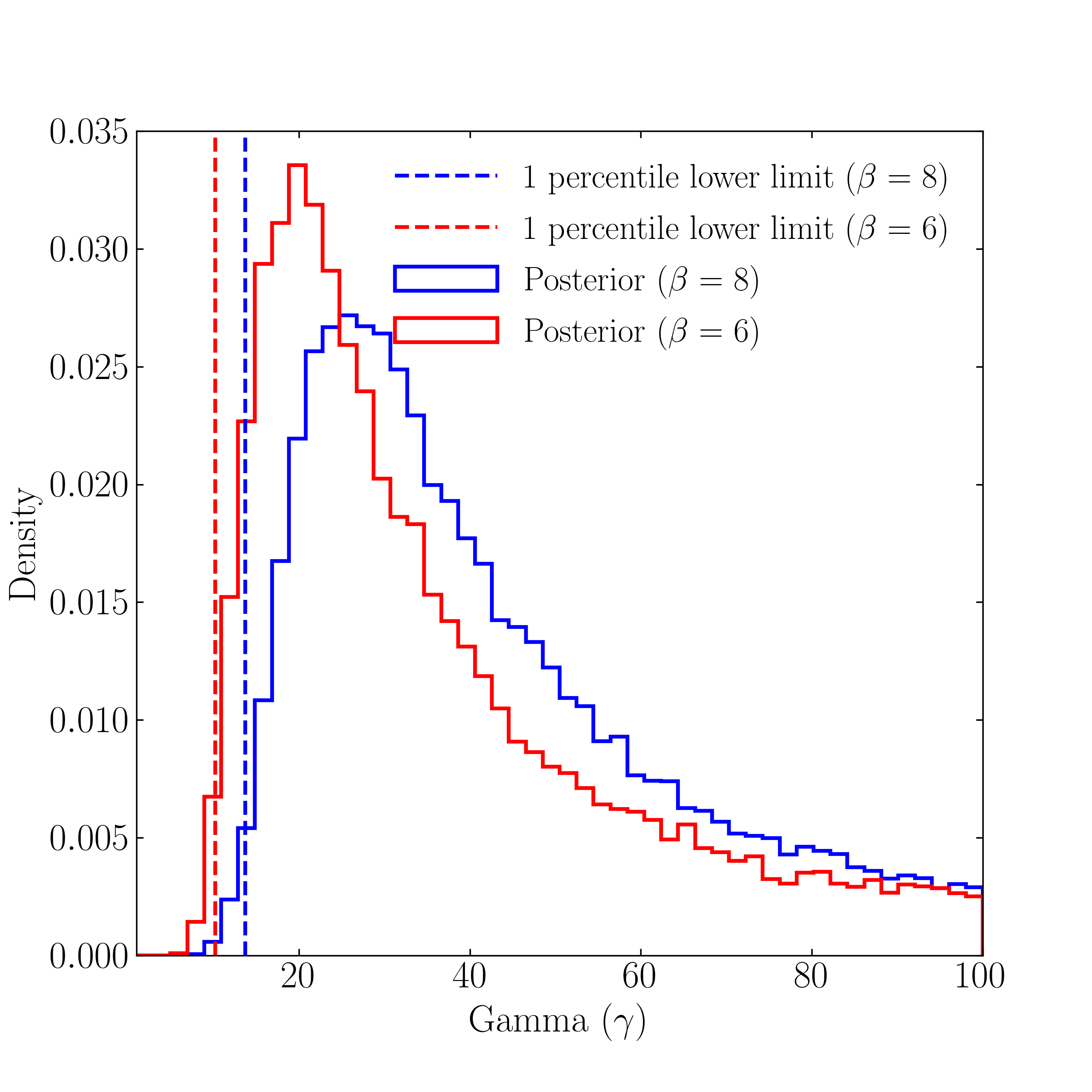}
  \caption{The posterior probability for $\gamma$ for a prior $\propto \gamma^{-2}$, and assuming $\beta=6$ and $\beta=8$.}
 \label{posterior}
\end{figure}

The choice of prior for $\gamma$ is not straightforward. Because we already know that a double Gaussian with the binaries offset by 0.75 mag. provides a satisfactory fit, this implies that $\gamma=\infty$ is consistent with the data, and all large values of $\gamma$ will be nearly equally preferred. This is problematic because adopting a uniform prior on $\gamma$ then leads to a meaningless posterior. Under this condition all that can usefully be achieved is to place a lower limit on $\gamma$. A pragmatic approach to this is to think in terms of an alternative parameter $\alpha=1/\gamma$, and to adopt a uniform prior in this parameter. This is in fact identical to adopting a prior on $\gamma$ of the form $\propto \gamma^{-2}$. This is the form we have chosen to adopt, with the range $\beta-1<\gamma<100$. We impose a lower limit for $\gamma$ of $\beta-1$ because $m(l)$ is undefined at $l=0$ for lower values of $\gamma$. In fact this lower limit does not come into play because the likelihood  drops off so steeply as a function of $\gamma$ well before the limit. The upper limit for $\gamma$ is arbitrary, and we discuss this point below. For the mass-luminosity relation we conservatively select $\beta=6$. This is conservative compared to $\beta=8$ in the sense that it leads to lower values of $\gamma$. \footnote{Concerning the chosen prior, an alternative might have been to adopt a uniform prior for a parameter $\alpha^\prime=\log(\gamma)$. This is the same as a prior on $\gamma$ of the form $\propto \gamma^{-1}$, which is less conservative.}

The results for the posterior on $\gamma$, determined by MCMC sampling, are plotted in Fig. \ref{posterior}. The median value for $\beta=6$ is $\gamma=29$  and for $\beta=8$ is $\gamma=36$. The best-fit value of $n_1$ is slightly lower at 0.84 than from the double Gaussian fit, while $\sigma_M$ is almost identical. Plotting the function it is essentially identical to the double Gaussian plotted in Fig. \ref{FigModelFit}. This is as expected, and the median value of $\gamma$ itself is not particularly interesting or meaningful \---\ it suffices to say that it is very large. More interestingly we find that $99\%$ of the posterior probability lies at $\gamma>10$ for the more conservative $\beta=6$ case. Similarly large values for the lower limit on $\gamma$ were obtained even when we changed substantially the upper limit of the prior and the functional form of the prior. We conclude that $\gamma>10$ for unresolved binaries in the spectral range M7 to M9.5, and this is the important result from this analysis. Since $M_J$ falls by 0.5 mag. per spectral type, the typical difference in spectral type between primary and secondary is calculated to be less than half a spectral type. In effect unresolved binaries of ultracool M dwarfs are identical twins. 

Could the steep measured value of $\gamma$ be because of a bias arising from how the sample of \citet{Ahmed19} was selected? If the secondary were of a different spectral type to the primary would the unresolved binary still be selected, or would it be rejected because of the peculiar colours? It is easy to see that there would be no bias from this effect, because the colours from M7 to L0 lie along a linear sequence over the bands $izYJHK$ used in the selection \citep{Skrzypek15,Skrzypek16}. This means that any binaries with combinations from within this sequence would have colours that lie on the colour sequence, and so would be selected. So there is no bias against unresolved binaries with small values of $\Delta M$.

Although this analysis provides a strong lower limit on $\gamma$, it only applies over a limited mass range. This is because of the steep mass-luminosity relation, meaning that this survey is not sensitive to the detection of brown dwarf companions. A small number of unresolved late-M $+$ brown dwarf binaries have been found (e.g. \citealp{Blake2008}, \citealp{Biller2006}, \citealp{Burgasser2012}, \citealp{Burgasser2015}). In principle the true mass ratio distribution could be bimodal, with an additional peak due to secondaries in the brown dwarf mass range.

\subsection{The binary fraction}
\label{binaryfraction}

Using the maximum-likelihood solution for n1 based on the $p(d)$ in section \ref{binarymass}, the fraction of binaries is $n_2=1-n_1=16.2\%$. This is the MF for unresolved systems, assuming triples and higher do not contribute significantly, as found by \citet{Winters2019}. This corresponds to 438 systems. However this neglects resolved binaries. The proportion of resolved binaries is expected to be very small \citep{2007ApJ...668..492A,Winters2019}. \citet{Winters2019} find a lognormal distribution of separations for their lowest mass bin ($0.075 - 0.15 M_{\odot}$) with a mean 7 AU. Given the distribution of distances of objects in our sample, we can compute the expected distribution of angular separations. The median seeing in the UKIDSS LAS data is 0.8\,arcsec \citep[e.g.][]{2007MNRAS.375..213W}, while in the SDSS data it is 1.3\,arcsec. Binaries with separations less than 0.8\,arcsec will have been counted in our volume-complete sample. We would expect to detect binaries with separations larger than 1.3\,arcsec as resolved. In the range 0.8\,arcsec to 1.3\,arcsec objects would be missed since the photometry (two single objects in UKIDSS, one merged object in SDSS) would be inconsistent. The predicted number of binaries with separations  between 0.8 and 1.3\,arcsec is five. We will neglect this minor source of incompleteness. The predicted number with separations greater than 1.3\,arcsec is just two.

To identify secondaries with separation greater than 1.3\,arcsec we have made a search around every object in our volume complete sample (which all have accurate distances) for objects within a projected separation of 1000\,AU in the parent sample of \citet{Ahmed19}. Because the distance limits of the parent sample are much greater than for the volume-complete sample, any stars in the spectral range M7 to M9.5 that are companions will be found, as can be seen by reference to the distance limits plotted in Fig. \ref{figExplainRegion}. We also searched in the sample of L and T dwarfs of \citet{Skrzypek16}, which extends the search to later types, with the same magnitude limits $13<J<17.5$. However the upper distance limit falls rapidly with spectral type so this search would not find all companions to primaries in the volume-complete sample. When a potential match was found within the specified angular separation limit, we then checked that the distances matched, using the {\em Gaia} parallax if available, otherwise the distance computed from the absolute magnitudes in Table \ref{tabVolSample}. A further visual confirmation was performed with SDSS DR13 images.  Using this method we identified four candidate wide binary pairs where both stars are in the volume complete sample. This increases the number of binaries by four, and at the same time reduces the total number of systems by four, to 2702. We additionally found five candidate wide binaries where the secondary is not in the volume complete sample. 

Combining the number of unresolved binaries with our candidate wide binaries, we measure 447 binary systems with an uncertainty of 21 (assuming Poisson statistics) from a total number of 2702 systems, yielding a total binary fraction, equal to the MF, of 16.5$\pm$0.8\%. This value is similar to but a little smaller than the values $20-22\%$ measured by \citet{2007ApJ...668..492A} and \citet{Winters2019}. Their samples are for spectral types M6 and later and M4 and later, respectively. The small difference is consistent with the trend of decreasing binary fraction with spectral type.
Interestingly our binary fraction of 16.5\% is close to the value found for L-dwarfs by \citealp{Reid2008}, suggesting that the M7-M9.5 dwarfs have more similar formation histories to the L-dwarfs than to the higher mass, early M-dwarfs.

\section{Summary}
\label{summary}

We have derived a volume-complete sample of 2706 M7.0-M9.5 dwarf systems, where the distance limits vary with $G-J$ colour, a proxy for luminosity. The sample benefits from accurate distances, and is unbiased with respect to multiplicity. We have determined the relation between $M_J$ and spectral type over this spectral range, measured the unresolved binary fraction, and determined the distribution of mass ratios of the stars in the unresolved binary systems. We have also measured the small fraction of resolved binaries where both the primary and the secondary are in this spectral range. The main results of our analysis of this sample are:
\begin{enumerate}
  \item We present a revised absolute magnitude - spectral type relation for the ultracool M-dwarfs. This revised relation relative to the determination of \citet{2012ApJS..201...19D} is on average 0.5 mag. brighter.  We present evidence that the differences are due to differences in the measured spectral types between the uniform BOSS sample of \citet{Schmidt2015} and the stars in the sample of  \citet{2012ApJS..201...19D}, with an average offset of close to one spectral subtype. \\
  \item We find that the distribution of mass ratios in unresolved binaries in this spectral range $f(q)\propto q^\gamma$ is very steep, with $\gamma > 10$ $(99\%$ probability). In effect unresolved binaries of ultracool M dwarfs are identical twins. \\
  \item We provide an estimate for the multiplicity fraction for M7 to M9.5 dwarfs of 16.5$\pm0.8\%$, consistent with previous estimates. Of these $98\%$ are unresolved. \\
  \item The spread in absolute magnitude for ultracool M dwarfs is $\sigma=0.21$\,mag. at fixed colour.
\end{enumerate}

\section*{Acknowledgements}

The authors wish to thank the anonymous referee for providing a helpful review which improved the clarity of the paper.
We are grateful to Daniel Mortlock for helpful discussions.
This work has made use of results from the European Space Agency (ESA) space mission \textit{Gaia},
the data from which were processes by the \textit{Gaia Data Processing and Analysis Consortium} (DPAC). Funding for the DPAC has been provided by national institutions, in particular the institutions participating in the \textit{Gaia} Multinational Agreement. The \textit{Gaia} mission website is \url{http://www.cosmos.esa.int/gaia}.
This work was supported by Grant ST/S505432/1 from the Science and Technology Facilities Council.

\section*{Data availability}

The data underlying this article will be shared on reasonable request to the corresponding author.



\bibliographystyle{mnras}
\bibliography{references} 

\begin{thebibliography}{}
\makeatletter
\relax
\def\mn@urlcharsother{\let\do\@makeother \do\$\do\&\do\#\do\^\do\_\do\%\do\~}
\def\mn@doi{\begingroup\mn@urlcharsother \@ifnextchar [ {\mn@doi@}
  {\mn@doi@[]}}
\def\mn@doi@[#1]#2{\def\@tempa{#1}\ifx\@tempa\@empty \href
  {http://dx.doi.org/#2} {doi:#2}\else \href {http://dx.doi.org/#2} {#1}\fi
  \endgroup}
\def\mn@eprint#1#2{\mn@eprint@#1:#2::\@nil}
\def\mn@eprint@arXiv#1{\href {http://arxiv.org/abs/#1} {{\tt arXiv:#1}}}
\def\mn@eprint@dblp#1{\href {http://dblp.uni-trier.de/rec/bibtex/#1.xml}
  {dblp:#1}}
\def\mn@eprint@#1:#2:#3:#4\@nil{\def\@tempa {#1}\def\@tempb {#2}\def\@tempc
  {#3}\ifx \@tempc \@empty \let \@tempc \@tempb \let \@tempb \@tempa \fi \ifx
  \@tempb \@empty \def\@tempb {arXiv}\fi \@ifundefined
  {mn@eprint@\@tempb}{\@tempb:\@tempc}{\expandafter \expandafter \csname
  mn@eprint@\@tempb\endcsname \expandafter{\@tempc}}}

\bibitem[\protect\citeauthoryear{{Ahmed} \& {Warren}}{{Ahmed} \&
  {Warren}}{2019}]{Ahmed19}
{Ahmed} S.,  {Warren} S.~J.,  2019, \mn@doi [\aap]
  {10.1051/0004-6361/201834591}, \href
  {https://ui.adsabs.harvard.edu/abs/2019A&A...623A.127A} {623, A127}

\bibitem[\protect\citeauthoryear{{Allen}}{{Allen}}{2007}]{2007ApJ...668..492A}
{Allen} P.~R.,  2007, \mn@doi [\apj] {10.1086/521207}, \href
  {https://ui.adsabs.harvard.edu/abs/2007ApJ...668..492A} {668, 492}

\bibitem[\protect\citeauthoryear{{Bailer-Jones}}{{Bailer-Jones}}{2015}]{Bailer-Jones2015}
{Bailer-Jones} C. A.~L.,  2015, \mn@doi [\pasp] {10.1086/683116}, \href
  {https://ui.adsabs.harvard.edu/abs/2015PASP..127..994B} {127, 994}

\bibitem[\protect\citeauthoryear{{Baraffe} \& {Chabrier}}{{Baraffe} \&
  {Chabrier}}{1996}]{BaraffeChabrierMLR}
{Baraffe} I.,  {Chabrier} G.,  1996, \mn@doi [\apjl] {10.1086/309988}, \href
  {https://ui.adsabs.harvard.edu/abs/1996ApJ...461L..51B} {461, L51}

\bibitem[\protect\citeauthoryear{{Bardalez Gagliuffi} et~al.,}{{Bardalez
  Gagliuffi} et~al.}{2019}]{Gagliuffi19}
{Bardalez Gagliuffi} D.~C.,  et~al., 2019, \mn@doi [\apj]
  {10.3847/1538-4357/ab253d}, \href
  {https://ui.adsabs.harvard.edu/abs/2019ApJ...883..205B} {883, 205}

\bibitem[\protect\citeauthoryear{{Biller}, {Kasper}, {Close}, {Brandner}  \&
  {Kellner}}{{Biller} et~al.}{2006}]{Biller2006}
{Biller} B.~A.,  {Kasper} M.,  {Close} L.~M.,  {Brandner} W.,   {Kellner} S.,
  2006, \mn@doi [\apjl] {10.1086/504256}, \href
  {https://ui.adsabs.harvard.edu/abs/2006ApJ...641L.141B} {641, L141}

\bibitem[\protect\citeauthoryear{{Binney} \& {Merrifield}}{{Binney} \&
  {Merrifield}}{1998}]{Binney98}
{Binney} J.,  {Merrifield} M.,  1998, {Galactic Astronomy}

\bibitem[\protect\citeauthoryear{{Blake}, {Charbonneau}, {White}, {Torres},
  {Marley}  \& {Saumon}}{{Blake} et~al.}{2008}]{Blake2008}
{Blake} C.~H.,  {Charbonneau} D.,  {White} R.~J.,  {Torres} G.,  {Marley}
  M.~S.,   {Saumon} D.,  2008, \mn@doi [\apjl] {10.1086/588754}, \href
  {https://ui.adsabs.harvard.edu/abs/2008ApJ...678L.125B} {678, L125}

\bibitem[\protect\citeauthoryear{{Burgasser}, {Kirkpatrick}, {Cruz}, {Reid},
  {Leggett}, {Liebert}, {Burrows}  \& {Brown}}{{Burgasser}
  et~al.}{2006}]{2006ApJS..166..585B}
{Burgasser} A.~J.,  {Kirkpatrick} J.~D.,  {Cruz} K.~L.,  {Reid} I.~N.,
  {Leggett} S.~K.,  {Liebert} J.,  {Burrows} A.,   {Brown} M.~E.,  2006,
  \mn@doi [\apjs] {10.1086/506327}, \href
  {https://ui.adsabs.harvard.edu/abs/2006ApJS..166..585B} {166, 585}

\bibitem[\protect\citeauthoryear{{Burgasser}, {Luk}, {Dhital}, {Bardalez
  Gagliuffi}, {Nicholls}, {Prato}, {West}  \& {L{\'e}pine}}{{Burgasser}
  et~al.}{2012}]{Burgasser2012}
{Burgasser} A.~J.,  {Luk} C.,  {Dhital} S.,  {Bardalez Gagliuffi} D.,
  {Nicholls} C.~P.,  {Prato} L.,  {West} A.~A.,   {L{\'e}pine} S.,  2012,
  \mn@doi [\apj] {10.1088/0004-637X/757/2/110}, \href
  {https://ui.adsabs.harvard.edu/abs/2012ApJ...757..110B} {757, 110}

\bibitem[\protect\citeauthoryear{{Burgasser} et~al.,}{{Burgasser}
  et~al.}{2015}]{Burgasser2015}
{Burgasser} A.~J.,  et~al., 2015, \mn@doi [\aj] {10.1088/0004-6256/149/3/104},
  \href {https://ui.adsabs.harvard.edu/abs/2015AJ....149..104B} {149, 104}

\bibitem[\protect\citeauthoryear{{Close}, {Siegler}, {Freed}  \&
  {Biller}}{{Close} et~al.}{2003}]{2003ApJ...587..407C}
{Close} L.~M.,  {Siegler} N.,  {Freed} M.,   {Biller} B.,  2003, \mn@doi [\apj]
  {10.1086/368177}, \href
  {https://ui.adsabs.harvard.edu/abs/2003ApJ...587..407C} {587, 407}

\bibitem[\protect\citeauthoryear{{Cruz} et~al.,}{{Cruz}
  et~al.}{2007}]{Cruz2007}
{Cruz} K.~L.,  et~al., 2007, \mn@doi [\aj] {10.1086/510132}, \href
  {https://ui.adsabs.harvard.edu/abs/2007AJ....133..439C} {133, 439}

\bibitem[\protect\citeauthoryear{{Dieterich}, {Henry}, {Golimowski}, {Krist}
  \& {Tanner}}{{Dieterich} et~al.}{2012}]{2012AJ....144...64D}
{Dieterich} S.~B.,  {Henry} T.~J.,  {Golimowski} D.~A.,  {Krist} J.~E.,
  {Tanner} A.~M.,  2012, \mn@doi [\aj] {10.1088/0004-6256/144/2/64}, \href
  {https://ui.adsabs.harvard.edu/abs/2012AJ....144...64D} {144, 64}

\bibitem[\protect\citeauthoryear{{Duch{\^e}ne} \& {Kraus}}{{Duch{\^e}ne} \&
  {Kraus}}{2013}]{2013ARA&A..51..269D}
{Duch{\^e}ne} G.,  {Kraus} A.,  2013, \mn@doi [\araa]
  {10.1146/annurev-astro-081710-102602}, \href
  {https://ui.adsabs.harvard.edu/abs/2013ARA&A..51..269D} {51, 269}

\bibitem[\protect\citeauthoryear{{Dupuy} \& {Liu}}{{Dupuy} \&
  {Liu}}{2012}]{2012ApJS..201...19D}
{Dupuy} T.~J.,  {Liu} M.~C.,  2012, \mn@doi [\apjs]
  {10.1088/0067-0049/201/2/19}, \href
  {https://ui.adsabs.harvard.edu/abs/2012ApJS..201...19D} {201, 19}

\bibitem[\protect\citeauthoryear{{Gaia Collaboration} et~al.,}{{Gaia
  Collaboration} et~al.}{2018}]{2018A&A...616A...1G}
{Gaia Collaboration} et~al., 2018, \mn@doi [\aap]
  {10.1051/0004-6361/201833051}, \href
  {https://ui.adsabs.harvard.edu/abs/2018A&A...616A...1G} {616, A1}

\bibitem[\protect\citeauthoryear{{Gizis}, {Reid}, {Knapp}, {Liebert},
  {Kirkpatrick}, {Koerner}  \& {Burgasser}}{{Gizis}
  et~al.}{2003}]{2003AJ....125.3302G}
{Gizis} J.~E.,  {Reid} I.~N.,  {Knapp} G.~R.,  {Liebert} J.,  {Kirkpatrick}
  J.~D.,  {Koerner} D.~W.,   {Burgasser} A.~J.,  2003, \mn@doi [\aj]
  {10.1086/374991}, \href
  {https://ui.adsabs.harvard.edu/abs/2003AJ....125.3302G} {125, 3302}

\bibitem[\protect\citeauthoryear{{Law}, {Hodgkin}  \& {Mackay}}{{Law}
  et~al.}{2008}]{2008MNRAS.384..150L}
{Law} N.~M.,  {Hodgkin} S.~T.,   {Mackay} C.~D.,  2008, \mn@doi [\mnras]
  {10.1111/j.1365-2966.2007.12675.x}, \href
  {https://ui.adsabs.harvard.edu/abs/2008MNRAS.384..150L} {384, 150}

\bibitem[\protect\citeauthoryear{{Lindegren} et~al.,}{{Lindegren}
  et~al.}{2018}]{2018A&A...616A...2L}
{Lindegren} L.,  et~al., 2018, \mn@doi [\aap] {10.1051/0004-6361/201832727},
  \href {https://ui.adsabs.harvard.edu/abs/2018A&A...616A...2L} {616, A2}

\bibitem[\protect\citeauthoryear{{Marshall}, {Tananbaum}, {Avni}  \&
  {Zamorani}}{{Marshall} et~al.}{1983}]{Marshall1983}
{Marshall} H.~L.,  {Tananbaum} H.,  {Avni} Y.,   {Zamorani} G.,  1983, \mn@doi
  [\apj] {10.1086/161016}, \href
  {https://ui.adsabs.harvard.edu/abs/1983ApJ...269...35M} {269, 35}

\bibitem[\protect\citeauthoryear{{Reid}, {Cruz}, {Burgasser}  \& {Liu}}{{Reid}
  et~al.}{2008}]{Reid2008}
{Reid} I.~N.,  {Cruz} K.~L.,  {Burgasser} A.~J.,   {Liu} M.~C.,  2008, \mn@doi
  [\aj] {10.1088/0004-6256/135/2/580}, \href
  {https://ui.adsabs.harvard.edu/abs/2008AJ....135..580R} {135, 580}

\bibitem[\protect\citeauthoryear{{Reyl{\'e}} et~al.,}{{Reyl{\'e}}
  et~al.}{2010}]{Reyle2010}
{Reyl{\'e}} C.,  et~al., 2010, \mn@doi [\aap] {10.1051/0004-6361/200913234},
  \href {https://ui.adsabs.harvard.edu/abs/2010A&A...522A.112R} {522, A112}

\bibitem[\protect\citeauthoryear{{Schmidt}, {Hawley}, {West}, {Bochanski},
  {Davenport}, {Ge}  \& {Schneider}}{{Schmidt} et~al.}{2015}]{Schmidt2015}
{Schmidt} S.~J.,  {Hawley} S.~L.,  {West} A.~A.,  {Bochanski} J.~J.,
  {Davenport} J. R.~A.,  {Ge} J.,   {Schneider} D.~P.,  2015, \mn@doi [\aj]
  {10.1088/0004-6256/149/5/158}, \href
  {https://ui.adsabs.harvard.edu/abs/2015AJ....149..158S} {149, 158}

\bibitem[\protect\citeauthoryear{{Skrzypek}, {Warren}, {Faherty}, {Mortlock},
  {Burgasser}  \& {Hewett}}{{Skrzypek} et~al.}{2015}]{Skrzypek15}
{Skrzypek} N.,  {Warren} S.~J.,  {Faherty} J.~K.,  {Mortlock} D.~J.,
  {Burgasser} A.~J.,   {Hewett} P.~C.,  2015, \mn@doi [\aap]
  {10.1051/0004-6361/201424570}, \href
  {https://ui.adsabs.harvard.edu/abs/2015A&A...574A..78S} {574, A78}

\bibitem[\protect\citeauthoryear{{Skrzypek}, {Warren}  \& {Faherty}}{{Skrzypek}
  et~al.}{2016}]{Skrzypek16}
{Skrzypek} N.,  {Warren} S.~J.,   {Faherty} J.~K.,  2016, \mn@doi [\aap]
  {10.1051/0004-6361/201527359}, \href
  {https://ui.adsabs.harvard.edu/abs/2016A&A...589A..49S} {589, A49}

\bibitem[\protect\citeauthoryear{{Ward-Duong} et~al.,}{{Ward-Duong}
  et~al.}{2015}]{Ward-Duong2015}
{Ward-Duong} K.,  et~al., 2015, \mn@doi [\mnras] {10.1093/mnras/stv384}, \href
  {https://ui.adsabs.harvard.edu/abs/2015MNRAS.449.2618W} {449, 2618}

\bibitem[\protect\citeauthoryear{{Warren} et~al.,}{{Warren}
  et~al.}{2007}]{2007MNRAS.375..213W}
{Warren} S.~J.,  et~al., 2007, \mn@doi [\mnras]
  {10.1111/j.1365-2966.2006.11284.x}, \href
  {https://ui.adsabs.harvard.edu/abs/2007MNRAS.375..213W} {375, 213}

\bibitem[\protect\citeauthoryear{{West} et~al.,}{{West}
  et~al.}{2011}]{West2011}
{West} A.~A.,  et~al., 2011, \mn@doi [\aj] {10.1088/0004-6256/141/3/97}, \href
  {https://ui.adsabs.harvard.edu/abs/2011AJ....141...97W} {141, 97}

\bibitem[\protect\citeauthoryear{{Winters} et~al.,}{{Winters}
  et~al.}{2019}]{Winters2019}
{Winters} J.~G.,  et~al., 2019, \mn@doi [\aj] {10.3847/1538-3881/ab05dc}, \href
  {https://ui.adsabs.harvard.edu/abs/2019AJ....157..216W} {157, 216}

\makeatother
\end{thebibliography}


\bsp	
\label{lastpage}
\end{document}